\documentstyle[eqsecnum,prb,aps,twocolumn,epsf]{revtex}
\newcommand{\be}{\begin{equation}}
\newcommand{\ee}{\end{equation}}
\newcommand{\ba}{\begin{eqnarray}}
\newcommand{\ea}{\end{eqnarray}}

\begin{document}    

\draft  

\title{Bosonization and effective vector-field theory of the
fractional quantum Hall effect}
 \author{K. Shizuya}
  \address{Yukawa Institute for Theoretical Physics\\
 Kyoto University,~Kyoto 606-8502,~Japan }

\maketitle

\begin{abstract} 
The electromagnetic characteristics of the fractional quantum
Hall states are studied by formulating an effective
vector-field theory that takes into account projection
to the exact Landau levels from the beginning.
The effective theory is constructed, via bosonization, from the
electromagnetic response of an incompressible and uniform state.
It does not refer to either the composite-boson or
composite-fermion picture, but properly reproduces the results
of the standard bosonic and fermionic Chern-Simons approaches,
thus revealing the universality of the long-wavelength
characteristics of the quantum Hall states and the
associated quasiparticles. In particular, the dual-field
Lagrangian of Lee and Zhang is obtained without invoking the
composite-boson picture. 
An argument is also given to verify, within a vector-field
version of the fermionic Chern-Simons theory, the identification
by Goldhaber and Jain of a composite fermion as a dressed
electron. 
\end{abstract}


\section{Introduction}
The fractional quantum Hall effect~\cite{TSG,Rev} (FQHE) results
from formation of incompressible quantum fluids that support
quasiparticles carrying fractional charges and statistics, as
all embodied in Laughlin's wave functions and his reasoning
resting on the gauge principle.~\cite{L}  
The early approaches based on variational wave
functions successfully  clarified some fundamental
aspects~\cite{L,H,Halp,GM,R,J} of the FQHE, which evolved into
the descriptions of the FQHE in terms of electron--flux
composites, composite bosons or composite fermions.

The bosonic~\cite{ZHK,LZ,Z} and fermionic~\cite{BW,LF,HLR}
Chern-Simons (CS) theories are the field-theoretical frameworks
that realize the composite-boson and composite-fermion pictures
of the FQHE and have been successful in describing the
long-wavelength characteristics of the fractional quantum
Hall (FQH) states. In the wave-function approaches it is crucial
to use wave functions projected to the lowest Landau level,
while no explicit account of such projection is taken in the
CS approaches. One might naturally wonder if and how the latter
are compatible with the Landau quantization such as Landau
levels and quenching of the electronic kinetic energy.

The purpose of this paper is to present a field-theoretical
approach to the FQH system, that takes account of the Landau
quantization from the very beginning.
The basic quantity we rely on is the electromagnetic response
of an incompressible and uniform state, which we calculate by handling
Landau-level mixing in a systematic way. We then use a
bosonization technique and construct from this response an
equivalent vector-field theory that describes the
electromagnetic characteristics of the FQH states and the
associated quasiparticles in the presence of the Coulomb
interaction.  This bosonic effective theory is derived without
recourse to either the composite-boson or composite-fermion
picture, but properly reproduces the random-phase-approximation
results of the bosonic and fermionic CS theories, thus
revealing the universality of the long-wavelength
characteristics of the FQH states. We examine some consequences
of it, especially for the composite fermions.

In Sec.~II we study the electromagnetic response of Hall
electrons by a method that enforces the Landau-level structure 
and gauge invariance of the Hall electron system. 
In Sec.~III we construct a bosonic
effective theory of the FQHE and discuss its consequences.  
In Sec.~IV we study a vector-field version of the fermionic CS
theory. Section~V is devoted to a summary and discussion.

\section{Electromagnetic response of Hall electrons}

Consider electrons confined to a plane with a perpendicular
magnetic field, described by the action:  
\begin{eqnarray}
S &=& \int dt d^{2}{\bf x}\, \psi^{\dag}({\bf x},t)\, 
(i\partial_{t}- H)\, \psi ({\bf x},t) +S_{\rm Coul}[\rho],
\label{Selectron}\\
H &=& {1\over{2M}} \left({\bf p}+e{\bf A}^{\!B}({\bf x})+e{\bf
A}({\bf x},t)\right)^{2} +eA_{0}({\bf x},t) . 
\label{Hone}
\end{eqnarray}
The vector potential ${\bf A}^{\!B}({\bf x})$ is taken to
supply  a uniform magnetic field $B_{z}=B>0$ normal to the
plane, and we make the Landau-gauge choice 
${\bf A}^{\!B}={1\over{2}}B\,(-y,0)$ below.
Our task in this section is to study the response of Hall
electrons to weak external potentials
$A_{\mu}({\bf x},t)=(A_{0}({\bf x},t),
{\bf A}({\bf x},t))$. 
[We suppose that $\mu$ runs over $(0,x,y)$ or $(0,1,2)$,
and denote ${\bf A}=(A_{x},A_{y})=(A_{1},A_{2})$, etc.
Remember that our $A_{0}$ equals minus the
conventional scalar potential.]  
For notational simplicity, we shall write 
$x=(t,{\bf x}),  d^{3}x = dt\,d^{2}{\bf x}, 
A_{\mu}(x) = A_{\mu}({\bf x},t)$, etc.,  when no confusion
arises; the electric charge
$e>0$ will also be suppressed by rescaling $eA_{\mu}\rightarrow
A_{\mu}$ in what follows.

The Coulomb interaction is a functional of the density
$\rho (x)=\psi^{\dag}(x)\psi(x)$,
\begin{eqnarray}
S_{\rm Coul}[\rho] &=&
-{1\over{2}}\, \int 
d^{3}x d^{2}{\bf x'}\, \delta \rho (x)\, V({\bf x-x'})\, 
\delta \rho (x') ,
\label{CI}
\end{eqnarray}
where $\delta \rho (x)=\rho (x)-\bar{\rho}$ stands for the
deviation from the average electron density $\bar{\rho}$.
It is convenient to write it in an equivalent form linear in 
$\rho(x)$ using the Hubbard-Stratonovich field $\chi (x)$, 
\begin{equation}
S'_{\rm Coul}[\rho] = \int d^{3}x \left[
{1\over{2}}\,\chi\, \Gamma [-i\nabla]\, \chi 
- \chi\,\delta \rho \right] , 
\label{linearizedCI} 
\end{equation}
where $\Gamma [{\bf p}] = 1/V[{\bf p}]$ in terms
of the Fourier transform of $V({\bf x})$. Note that as for the
electron sector the effect of this linearization is to replace
$A_{0}(x)$ by  $A_{0}(x)+\chi (x)$ in the Hamiltonian $H$ of
Eq.~(\ref{Hone}). Accordingly, in the rest of this section 
we shall denote $A_{0}+\chi$ as $A_{0}$ and focus on the
one-body part of $S$.  Explicit account of the Coulomb
interaction will be taken in Secs.~III and IV again.

The eigenstates of $H$ with $A_{\mu}=0$ are Landau
levels $|N\rangle = |n, y_{0}\rangle$ of energy $\omega
(n+{\scriptstyle {1\over2}})$, labeled by integers 
$n = 0,1,2,\cdots$, and $y_{0}=\ell^{2}\,p_{x}$, where
$\omega \equiv eB/M$ is the cyclotron energy 
and ${\ell}\equiv 1/\sqrt{eB}$ is the magnetic length.
The external potentials $A_{\mu}(x)$ modify this
level structure, and the desired response of the electron is
obtained by diagonalizing the Hamiltonian $H$ with respect to
the true Landau levels $\{n\}$ (or, projecting $H$ into the true
levels).

For actual calculations it is convenient to pass to the
$|N\rangle$ representation via a unitary transformation
$\langle {\bf x}| N\rangle$, with the expansion  
$\psi ({\bf x},t)=\sum_{N} \psi_{N}(t) 
\langle {\bf x}| N\rangle$. The
translation is simple:~\cite{KSw} The relative coordinates
$(y-y_{0})/\ell$ and $\ell\,p_{y}$ turn into the matrices $Y$
and $P$ of a harmonic oscillator with
$[Y,P]=i1$, and the arguments ${\bf x}=(x,y)$ of
$A_{\mu}({\bf x},t)$ are replaced by 
the operators $\hat{{\bf x}}=(\hat{x},\hat{y})$,
\begin{equation}
 \hat{x} \equiv x_{0} + \ell P , \ \ 
\hat{y}\equiv  y_{0}+ \ell Y ,
   \label{xy}
\end{equation}
with $[\hat{x},\hat{y}]=0$. Here $y_{0}$ and
$x_{0}\equiv i\ell^{2}\partial /\partial y_{0}$ 
stand for the center coordinates of an orbiting electron
with uncertainty
$[x_{0}, y_{0}]=i\ell^{2}$. 
The action is thereby rewritten as 
\begin{eqnarray}
 S &=& \int\! dt\, d y_{0}\! \sum_{m,n=0}^{\infty}\!
\psi^{\dag}_{m}(y_{0},t) \left(i\delta_{m n} \partial_{t}
-{\hat H}_{m n} \right) \psi_{n}(y_{0},t),  \nonumber \\
\hat{H} &=& \omega \left\{ 
[\bar{Z} -i\ell\bar{A}(\hat{\bf x},t)] 
[Z + i\ell\, A(\hat{\bf x},t)] + {\scriptstyle {1\over2}}
\right\} 
\nonumber\\&&
+{1\over{2M}} h_{z}(\hat{\bf x},t) + A_{0}(\hat{\bf x},t),
\label{hatH}
\end{eqnarray}
with
\begin{eqnarray}
 A(x) &=& \left\{A_{y}(x) +
iA_{x}(x)\right\}/\sqrt{2},\ \bar{A}(x) = A(x)^{\dag};
\nonumber\\
 h_{z}(x)& =&\partial_{x}A_{y}(x)-\partial_{y}A_{x}(x).
\end{eqnarray} 
In the above we have set
$\psi_{N}(t)\rightarrow \psi_{n}(y_{0},t)$ and defined 
$Z=(Y +iP)/\sqrt{2}$ and $\bar{Z} = Z^{\dag}$ so that 
$Z_{mn}=\sqrt{n}\,\delta_{m,n-1}$ and $[Z,\bar{Z}]=1$.
In this section we use the Landau gauge but all the
manipulations are carried over to the symmetric gauge as
well.

The Hamiltonian $\hat{H}$ is an infinite-dimensional matrix
in Landau-level indices and an operator in $y_{0}$ and
$x_{0}=i\ell^{2}\,\partial_{y_{0}}$. It has a gauge symmetry
far larger than the electromagnetic gauge invariance:
The change of bases in $N$ space,
$|n,y_{0}\rangle \rightarrow|n',y'_{0}\rangle$, 
induces a unitary transformation
$G=\{G_{mn}(x_{0},y_{0},t)\}$ of the field operator
\begin{equation}
\psi^{G}_{m}(y_{0},t)= \sum_{n=0}^{\infty}
G_{mn}(x_{0},y_{0},t)\,\psi_{n}(y_{0},t).
\end{equation}
The potentials $A_{\mu}$ thereby undergo the transformation
\begin{eqnarray}
A^{G}&=& G A(\hat{\bf x},t) G^{-1} 
- (i/\ell)\, [G,Z]G^{-1}, \nonumber\\  
A_{0}^{G}&=& G A_{0}(\hat{\bf x},t) G^{-1} 
- i\,G\, \partial_{t}G^{-1}, 
\label{AG}
\end{eqnarray}
which leaves the action $S$ invariant. 
Let us write 
\begin{equation}
G=\exp[i\ell\,\eta],
\end{equation}
where $\eta$ takes on values in the $U(\infty)$ or
$W_{\infty}$ algebra
\begin{equation}
\eta=\sum_{r,s=0}^{\infty}\eta_{rs}(x_{0},y_{0})
\bar{Z}^{r}Z^{s}
\label{Uinfinity}
\end{equation}
with operator-valued coefficients $\eta_{rs}(x_{0},y_{0})$. 
This $W_{\infty}$ transformation $G$ in general mixes
Landau levels. 
The original electromagnetic gauge invariance is
realized when $\eta$ is written in terms of 
$(\hat{x},\hat{y})$, i.e., of the special form
$\eta[\hat{x},\hat{y},t]$.

The reason for displaying the symmetry structure underlying the
Hall system is that it provides a powerful tool
to systematically project the Hamiltonian into the exact 
Landau levels for weak fields $|A_{\mu}|\ll \omega$. 
One may simply adjust $\eta$ so that
$\hat{H}^{G}=G(\hat{H} - i\partial_{t})G^{-1}$ is diagonal in
level indices. Some calculations along this line were given
earlier.~\cite{KSw} Here we demonstrate another advantage of
the method that, with a suitable transformation, the projection
is done in a manifestly gauge-covariant manner.  The
calculation itself, however, is rather independent of the main
line of our discussion and is left for Appendix A.
Here we quote only the result: 
The projected Hamiltonian, to $O(A^{2})$ and $O(\partial^{2})$,
reads
\begin{eqnarray}
\hat{H}^{G}&=& (\bar{Z}Z + {1\over2}) 
\Big\{\omega + {1\over{M}} (h_{z} +
\ell^{2}h_{z}^{2}) +{1\over2}\,\ell^{2}\,{\bf
\nabla\!\cdot\! E} \Big\}
\nonumber\\&&
+A_{0} -{\ell^{2}\over{2\omega}}\,{\bf E}^{2} 
+ {1\over2} \,\ell^{2}\,\epsilon^{\mu \nu
\lambda }A_{\mu}\partial_{\nu } A_{\lambda } + \cdots,
\label{HGprime}
\end{eqnarray}
where all the fields depend on $(x_{0},y_{0},t)$ and total
derivatives have been suppressed;
${\bf E}= -\partial_{0}{\bf A}+\nabla A_{0}$, and
$\epsilon^{\mu\nu\rho}$ is a totally-antisymmetric tensor with
$\epsilon^{012} =1$.

Since the Hamiltonian $\hat{H}^{G}$ is diagonal
in $n$, let us denote 
$(\hat{H}^{G})_{nn}=(n+{1\over2})\,\omega+V_{n}(x_{0},y_{0},t)$
and focus on a single level $n$. 
The cyclotron energy is now quenched and $V_{n}$ describes the
response of the Hall electron.  Let us suppose that in the
absence of external perturbations ($A_{\mu}=0$) a nondegenerate 
many-body state of uniform density is realized.  
Here we have in mind the
integer and fractional quantum Hall states and understand that
the Coulomb interaction is responsible for their formation.  
For such incompressible and uniform states it is not difficult to
translate 
$V_{n}(x_{0},y_{0},t)$ into the effective action or the
partition function in ${\bf x}$ space.  
To this end, replace first 
$(x_{0},y_{0}) \rightarrow (\hat{x},\hat{y})$ in $V_{n}$ to
form 
$\hat{V}_{n}\equiv V_{n}(\hat{x},\hat{y},t)$, and note that
$\hat{V}_{n}$ differs from $V_{n}$ by total divergences,
which will soon turn out irrelevant in the end. Then go back
to the ${\bf x}$ space with $\hat{V}_{n}$.
This yields an interaction term of the form
$\psi^{\dag}V_{n}(x,y,t)\psi$, which, for an electron state of
uniform density $\rho_{n}$ in the $n$th level, gives rise to
the effective Lagrangian 
$L_{n}^{\rm eff}= -\rho_{n} V_{n}(x,y,t)$, or
\begin{eqnarray}
L_{n}^{\rm eff} = \rho_{n} \Big\{&&
-e A_{0}+{e^{2}\ell^{2}\over{2\omega}}\,{\bf E}^{2} 
- (n+{\scriptstyle {1\over2}})
{e^{2}\ell^{2}\over{M}}\,h_{z}^{2}
\nonumber\\&&
 - {1\over2} \,e^{2}\ell^{2}\,
\epsilon^{\mu \nu \lambda}A_{\mu}\partial_{\nu } A_{\lambda } 
+\cdots 
\Big\}
\label{LeffA}
\end{eqnarray} 
with the electric charge $e$  restored. Here all the
fields are functions of $(x,y,t)$ and total divergences
have been dropped. The $L_{n}^{\rm eff}$ summarizes the
$O(V_{n})$ electromagnetic response of the incompressible and uniform
state. It correctly reproduces earlier results~\cite{LF,sakita} 
for integer filling $\rho_{n} \rightarrow 1/(2\pi\ell^{2})$.
It is clear from our derivation that such a response is
determined uniquely for an incompressible state of general $\rho_{n}$,
independent of the detail of how it is formed; 
this is the key observation that enables one to use this response for
the discussion of the FQHE.

\section{bosonization and an effective theory}

The action $S=S[\psi,\psi^{\dag},A_{\mu}]$ in
Eq.~(\ref{Selectron}) describes the electron field $\psi (x)$
minimally coupled to the external potentials
$A_{\mu}(x)=(A_{0}(x),A_{k}(x))$, where $k$ runs over
$(1,2)=(x,y)$. In this section we study this electron
system from somewhat different angles and
construct a bosonic effective theory of the FQHE. 
The electromagnetic response of the system is summarized
in the partition function written as a functional integral 
\begin{equation}
W[A] = \int [d\psi][d\psi^{\dag}]\, 
e^{i S[\psi,\psi^{\dag}, A_{\mu}]} .
\label{Wa}
\end{equation}
[From now on $A$ simply refers to $A_{\mu}$, and not to 
$(A_{y} + iA_{x})/\sqrt{2}$ any more.]
The $W[A]$ encodes in its $A_{\mu}$ dependence the quantum
effect of the electron field $\psi$.
Once such $W[A]$ is known it is possible to reconstruct
it through the quantum fluctuations of a boson field. 
Let us first briefly review this procedure, known as functional
bosonization.~\cite{FS,schaposnik,BOS,BO}

The gauge invariance of the action 
$S[\psi,\psi^{\dag}, A_{\mu}] =
S[e^{-i\xi}\psi,\psi^{\dag}e^{i\xi}, A_{\mu} + 
\partial_{\mu}\xi]$ dictates that $W[A]$ is gauge-invariant,
i.e., $W[A_{\mu}]=W[A_{\mu}+\partial_{\mu}\xi]$, where 
$\partial_{\mu}=(\partial_{0},\partial_{1},\partial_{2})
\equiv (\partial_{t},\partial_{x},\partial_{y})$.
Thus integrating $W[A_{\mu}]$ over $\xi$ amounts to
an inessential change in the overall normalization,
yielding 
$W[A_{\mu}]=\int [d\xi]\, W[A_{\mu}+\partial_{\mu}\xi]$. 
Now let us rewrite the integral $\int [d\xi]$ as an
integral over a 3-vector field 
$v_{\mu}(x) =(v_{0},v_{1},v_{2})$, 
\begin{equation}
W[A] = \int [dv_{\lambda}] \, \delta (\epsilon^{\mu \nu
\rho} \partial_{\nu}v_{\rho})\, W[A+ v] ,
\end{equation}
where  the delta functional enforces the pure-gauge nature 
$v_{\mu} \sim \partial_{\mu}\xi$.
One can disentangle the pure-gauge constraint by making
use of a functional Fourier transform  with another
3-vector field $b_{\mu}=(b_{0},b_{1},b_{2})$.
A subsequent shift 
$v_{\mu}\rightarrow v_{\mu}-A_{\mu}$ in the functional
integral then yields the representation 
\begin{eqnarray}
W[A]&=&\int [db_{\lambda}]\,e^{ -i\int d^{3}x\,  
(A_{\mu} \epsilon^{\mu \nu\rho} \partial_{\nu}b_{\rho} ) +
iZ[b]}, \label{intdb}\\
e^{iZ[b]}&\equiv& \int [dv_{\lambda}]\,e^{ i\int
d^{3}x\,   (b_{\mu} \epsilon^{\mu \nu\rho}
\partial_{\nu}v_{\rho} ) }\, W[v].
\label{Seffb}
\end{eqnarray}
Here $Z[b]$ is obtained by Fourier transforming (given)
$W[v]$.  Note that the electron current 
$j_{\mu}=\delta S/\delta A_{\mu}$ is written as a rotation
$\epsilon^{\mu\nu\rho}\partial_{\nu}b_{\rho}$ in the $b$-boson
theory.

An advantage of the bosonic theory is that the bosonization
of the current holds exactly for the Coulomb
interaction~(\ref{CI}).  
To see this write the Coulomb interaction in linearized form
$S'_{\rm Coul}[\rho]$ and follow the bosonization procedure
(with a shift
$v_{\mu}\rightarrow v_{\mu}-A_{\mu} -\delta_{0\mu}\,\chi$).
Eliminating (or integrating over) the $\chi$ field then
entails the replacement 
$\rho \rightarrow \epsilon^{0ij}\partial_{i}b_{j}
=\partial_{1}b_{2}-\partial_{2}b_{1}\equiv b_{12}$ 
in $S_{\rm Coul}[\rho]$, and the $b$-boson theory is
described by the action 
\begin{equation}
S_{\rm eff}[b]=\int d^{3}x\,  
(-A \epsilon \partial b) + Z[b] + S_{\rm Coul}[b_{12}],
\label{Scb}
\end{equation}
where $Z[b]$ is now calculated from the one-body part of the action $S$
in Eq.~(\ref{Selectron}),
i.e., from
the $v$-field theory with the action 
\begin{equation}
S_{v}=\int d^{3}x\, b \epsilon \partial v  
-i {\rm Tr}\log(i\partial_{t} - H[v]).
\label{Sc}
\end{equation}
Here we have introduced compact notation
$A \epsilon \partial b \equiv
A_{\mu}\epsilon^{\mu \nu\rho} \partial_{\nu}b_{\rho}$, etc.,
which will be used from now on.

Actually the second term of Eq.~(\ref{Sc}) has been calculated
in the preceding section. Suppose that a nondegenerate many-body 
state of uniform density
$\bar{\rho}= \nu/(2\pi\ell^{2})$ is formed via the Coulomb
interaction (for $A_{\mu}=0$); take, for generality,
$n<\nu<n+1$, so that the lower $n+1$ Landau levels are
filled up.  Its response to weak electromagnetic potentials
$A_{\mu}(x)$ is obtained by collecting Eq.~(\ref{LeffA}) for
the filled levels, yielding 
$\log W[A] = i\int d^{3}x\, L[A;\nu]$ with  
\begin{eqnarray}
 L[A;\nu]= \bar{\rho}&&\Big[-A_{0}
-s_{B}\,\ell^{2}\,{1\over2}\,A\epsilon \partial A
\nonumber\\&&
+{\ell^{2}\over{2\omega}} (A_{k0})^{2} -
{\sigma(\nu)\,\ell^{2}\over{2M}}  (A_{12})^{2}\Big]  +
O(\partial^{3}),
\label{LAnu}
\end{eqnarray}
where
$A_{\mu\nu}\equiv \partial_{\mu}A_{\nu}-\partial_{\nu}A_{\mu}$ 
(so that $A_{k0}=E_{k}$ and $A_{12}=h_{z}$). Here
$\sigma(\nu)=1 + 2n - n(n+1)/\nu$ at each integer interval
$n<\nu\le n+1$; the modification needed for realistic
spin-resolved levels is obvious. 
For later use we have generalized $L[A;\nu]$ to accommodate the
sign $s_{B}\equiv {\rm sign}(B)=\pm1$ of $B_{z}=B$; 
remember that, when the magnetic field is reversed in
direction, the CS term $\propto A \epsilon \partial A$ (or the
Hall conductance) changes in sign.

Let us now substitute this $L[A;\nu]$ into Eq.~(\ref{Seffb})
and construct an effective theory of the $b_{\mu}$ field that
recovers it.
The exponent $\propto b \epsilon \partial v + L[v;\nu]$
is essentially quadratic in
$v_{\mu}$, and functional integration over $v_{\mu}$ is
carried out exactly. Here it is necessary to fix a gauge.
Fortunately there is a way to avoid such gauge-fixing
complications, that works in the presence of CS couplings, as
we explain below:  First make a shift 
$v_{\mu}=v'_{\mu} +f_{\mu}[b]$ and choose
$f_{\mu}[b]$ so that no direct coupling between
$v'_{\mu}$ and $b_{\mu}$ remains in the exponent,
which thereby is split into two terms $L[v';\nu]+L^{(0)}[b]$.
The integration over $v'_{\mu}$ requires fixing a gauge but is
done trivially, leaving no dependence on $b_{\mu}$.
All the dependence on $b_{\mu}$ is now isolated in the background
piece $f_{\mu }= (1/\bar{\rho}\ell^{2})\, s_{B}\,b_{\mu}
+ O(\partial\,b)$ and the desired  bosonic action is given by
$Z[b]=\int d^{3}x\,L^{(0)}[b]$ with 
\begin{eqnarray}
L^{(0)}[b]= &&- s_{B}\,{1\over{\ell^{2}}}\,b_{0}+
s_{B}\,{\pi\over{\nu}}\,b_{\mu} \epsilon^{\mu\nu\lambda}
\partial_{\nu}  b_{\lambda} 
\nonumber\\&&
+{\pi\over{\nu\,\omega}}\,(b_{k0})^{2} - 
{\sigma(\nu)\,\pi\over{\nu\,M}}\,(b_{12})^{2}  
+ O(\partial^{3}) ,
\label{Lzerob}
\end{eqnarray}
where the filling factor $\nu= 2\pi \ell^{2}\bar{\rho}$.
See Appendix B for details.
An effective action analogous to the above was discussed for
integer filling earlier.~\cite{BO}
In connection with the $v'_{\mu}$ integration we remark that any
simple choice of gauge for $v'_{\mu}$, e.g., 
$\partial_{k}v'_{k}=0$, takes a rather unconventional form 
$\partial_{k}v_{k}=\partial_{k}f_{k}[b]$ in terms of
$v_{\mu}$.  Still this is a perfectly legitimate choice, and it
has the advantage of achieving separation of the background
($b_{\mu}$ here) dependent piece at the classical level.

The bosonic theory allows one to handle the Coulomb interaction
exactly, as already remarked. It further admits inclusion of new
degrees of freedom, vortices, which describe quasiparticle
excitations over the FQH states.

Vortices arise only around special filling fractions,
at which nondegenerate many-body ground states are realized, as
revealed by Laughlin's reasoning~\cite{L} of
arriving at an excited state by adiabatically piercing an exact
many-body state with a thin solenoid of a unit flux quantum
$\phi_{D}=2\pi\hbar/e$; recall that the key element there
is the nondegeneracy of the ground state. The vortices
reside in a topologically nontrivial component of the phase of
the electron field.~\cite{L,LZ} Let us isolate 
the nontrivial portion by writing
$\psi(x) = e^{i\Theta (x)}\psi'(x)$ with 
$\epsilon^{0ij} \partial_{i} \partial_{j}\Theta (x) \not=0$,
where $\psi'(x)$ describes the  electrons  away from the
vortices so that they locally belong to the ground state with
uniform density $\bar{\rho}$. For vortices of
vorticity
$\{q_{i}\}$ and position $\{{\bf x}^{(i)}(t)\}$ the vortex
3-current $\tilde{j}_{\mu}=(\tilde{\rho},\tilde{j}_{k})=
(1/2\pi)\,\epsilon^{\mu\nu\rho}\partial_{\nu}\partial_{\rho}
\Theta(\{{\bf x}^{(i)}(t)\})$ is
written as 
\begin{eqnarray}
\tilde{j}_{\mu}(x)= \sum_{i} 
(1,\partial_{t}{\bf x}^{(i)}(t) )\,  
q_{i}\,\delta ({\bf x} -{\bf x}^{(i)}(t)) .
\end{eqnarray}

The partition function (\ref{Wa}), rewritten in terms of
$\psi'(x)$, is given by 
$W[A+\partial\, \Theta]$, which upon bosonization brings about 
the change 
$A\epsilon \partial b \rightarrow 
A\epsilon \partial b +2\pi \tilde{j}_{\mu}b_{\mu}$ 
in $S_{\rm eff}[b]$.
Thus the bosonic theory that takes into account
both the Coulomb interaction and vortices is described by the
Lagrangian
\begin{equation}
L_{\rm eff}[b]= -A_{\mu} \epsilon^{\mu\nu\rho}
\partial_{\nu}b_{\rho} 
-2\pi \tilde{j}_{\mu}b_{\mu}  + L^{(0)}[b] 
-{1\over{2}}\, \delta b_{12}\,V\, \delta b_{12} ,
\label{Leffb}
\end{equation}
where $\delta b_{12}\,V\, \delta b_{12} \equiv 
\int d^{2}{\bf y}\, \delta b_{12}(x)\,
V({\bf x -y})\, \delta b_{12}(y)$ for short and
$\delta b_{12}(x)= b_{12}(x) -\bar{\rho}$. 
Upon quantization the CS term $b\epsilon\partial b$ in
$L^{(0)}[b]$ combines with the kinetic term $(b_{k0})^{2}$ to
yield a ``mass'' gap $\omega$ while the $(b_{12})^{2}$ term
yields only a tiny $O(\nabla^{2})$ correction to it, as we shall
see soon.

Note that the $A \epsilon \partial b$ term combines
with the $(1/\ell^{2})b_{0}$ term in $L^{(0)}[b]$ to promote
$A_{\mu}$ to the full vector potential $A_{\mu}+ A^{B}_{\mu}$.
As a result, $L_{\rm eff}[b]$ almost precisely agrees with the
dual-field Lagrangian of Lee and Zhang~\cite{LZ} (LZ),
describing the low-energy features of the FQHE. 
The only difference to $O(\partial^{2})$ lies in the 
$(b_{12})^{2}$ term, which, however, is unimportant at long
wavelengths and actually negligible compared with the
$(b_{k0})^{2}$ term, as indicated by the ratio
$\omega/M \sim 10^{-7}$ typically.
We have thus practically reproduced the dual Lagrangian of LZ.

The LZ approach relies on the Chern-Simons-Landau-Ginzburg
(CSLG) theory realizing the composite-boson description of the
FQHE; there the quantum fluctuations of the composite-boson
field around the mean field $\rho \sim \bar{\rho}$ are converted,
via the dual transformation of Lee and Fisher~\cite{LeeF}, into
those of a 3-vector field $b_{\mu}^{\rm LZ}$.  
The $b_{\mu}^{\rm LZ}$ and our $b_{\mu}$ bosonize the electron
current in formally the same way. In the CSLG theory no account
of the Landau-level structure is taken, and the proper
electromagnetic responses of the FQH states are obtained through
the random-phase approximation (RPA) beyond mean field.

In contrast, our approach takes explicit account of Landau levels
and relies on the electromagnetic response of a uniform-density
state, which, via bosonization, is transformed into the dynamics
of the $b$ field to study the long-wavelength characteristics of
the FQH states.

Thus, the fact that we have arrived at the LZ dual Lagrangian 
without invoking the composite-boson picture would reveal the
following: 
(1) The long-wavelength electromagnetic responses of
the FQH states, as governed by the LZ dual Lagrangian, are
determined universally by the filling factor and some
single-electron characteristics, independent of the details of the
FQH states.  
(2) In the CSLG theory the RPA properly recovers the effect of the
Landau-level structure, crucial for determining the
electromagnetic response.

Having established the connection to the composite-boson approach,
let us now derive the electromagnetic response starting
from the bosonic Lagrangian $L_{\rm eff}[b]$.
As before the $b$ integration is done by a suitable shift 
$b_{\mu}=b'_{\mu} +f_{\mu}$, with the result (in obvious
compact notation)
\begin{eqnarray}
 L_{b}[A;\nu]&=& 
\bar{\rho}\,\ell^{2}\,\Big[ -{1\over{\ell^{2}}}\,A_{0}
-s_{B}\,{1\over2}\,A {1\over{D}}\,\epsilon \partial A 
\nonumber\\&&
 +{1\over{2\omega}} A_{k0} {1\over{D}} A_{k0} 
-{\nu\over{4\pi}}  A_{12}{1\over{D}} \Sigma A_{12}
\Big] ,
\label{LArpa}
\end{eqnarray}
where
\begin{equation}
\Sigma = V + {2\pi\sigma(\nu)\over{\nu M}}, \ 
D=1+{1\over{\omega^{2}}}\,\partial_{0}^{2} - {\nu
\over{2\pi\omega}}\,\Sigma\, \nabla^{2} ;
\label{SigmaD}
\end{equation}
see Appendix B for details. 
This $L_{b}[A;\nu]$ improves the original response
$L[A;\nu]$ in Eq.~(\ref{LAnu}) and agrees with the RPA
result in the composite-boson theory.~\cite{LZ,Z}  
One can read off, e.g., the density-density correlation function
from the $A_{0}A_{0}$ portion of the $A_{k0}D^{-1}A_{k0}$ term.
One also learns from $D$ that the Coulomb interaction
modifies the dispersion of the cyclotron mode so that 
\begin{eqnarray}
\omega ({\bf p}) &=& \omega + {1\over{2}}\,
\bar{\rho}\ell^{2}\, {\bf p}^{2}\,
\left( V[{\bf p}]+ {2\pi\sigma(\nu)\over{\nu M}} \right)
\nonumber\\
&\approx& \omega +  
{1\over{2}}\,\bar{\rho}\ell^{2}\,{\bf p}^{2}\,V[{\bf p}]
\label{omegap}
\end{eqnarray}
in accordance with Kohn's theorem.~\cite{LZ,K} Here we see
explicitly that the $(b_{12})^{2}$ terms in $L_{\rm eff}[b]$
yields only a tiny nonleading correction.

Observe here that, in spite of the quenched electronic kinetic
energy, the $b_{\mu}$ field correctly acquires
a dispersion of the Landau gap $\sim\omega$ (through level
mixing caused by electromagnetic couplings). The cyclotron mode
is readily identified by first isolating a mean-field piece from
the quantum component $b'_{\mu}$ so that 
$b'_{\mu}=\delta b'_{\mu} +\langle b'\rangle_{\mu}$ with
$\langle b'\rangle_{12}=\bar{\rho}$.
In the Coulomb gauge $\partial_{k}\delta b'_{k}=0$ one
can write $\delta
b'_{i}=\epsilon^{0ij}(\partial_{j}/\sqrt{-\nabla^{2}}) \, \zeta$
and show that $\zeta$ is the canonical scalar field with 
the dispersion $\omega({\bf p})$.

In Eq.~(\ref{LArpa}) we have set $\tilde{j}_{\mu}=0$ for
simplicity, but the vortices are easily recovered by
the substitution
$A_{\mu} \rightarrow A_{\mu}+ \partial_{\mu}\Theta$ and 
$A_{\mu\nu}\rightarrow A_{\mu\nu}
+2\pi\,\epsilon^{\mu\nu\lambda}\tilde{j}_{\lambda}$.
The dynamics of vortices is immediately read from 
$L_{b}[A_{\mu} + \partial_{\mu}\Theta;\nu]$. 
In particular, the $A \epsilon \partial A$ term contains a
vortex coupling like 
$-\nu\,\tilde{j}_{\mu}A_{\mu}$, which shows that a vortex of
vorticity $q$ has charge $-\nu q e$. 
(We have set $s_{B}=1$ here.)

It is possible to read off the vortex charge directly from
$L_{\rm eff}[b]$ in Eq.~(\ref{Leffb}). 
There the vortex enters only through the
$-2\pi \tilde{j}_{\mu}\, b_{\mu}$ coupling and no direct
coupling to $A_{\mu}$ exists.
As remarked in deriving Eq.~(\ref{Lzerob}), however,
the $b_{\mu}$ field acquires, in the presence of the
electromagnetic coupling
$-A_{\mu} \epsilon^{\mu \nu\rho} \partial_{\nu}b_{\rho}$, 
a background component, i.e.,
$b_{\mu}=b'_{\mu} +f_{\mu}$ with 
$f_{\mu}= s_{B}\,(\nu/2\pi)\, A_{\mu} + O(\partial A)$, 
where the coefficient $\nu/2\pi$ derives from 
the $b\epsilon \partial b$ term. Thus the vortex is coupled to
$A_{\mu}$ through the background piece
\begin{equation}
-2\pi \tilde{j}_{\mu}\,b_{\mu}= - \nu\, 
\tilde{j}_{\mu}A_{\mu} + \cdots,
\end{equation}
which reveals the vortex charge of $-\nu q e$.

\section{Composite fermions}

The composite fermions are electrons carrying an even number of
flux quanta. In the CS approach the electron field
$\psi(x)$  is converted to the composite-fermion field
$\phi(x)$ by a singular gauge transformation that attaches an
even number $\alpha=2p$ of flux quanta $\phi_{D}=2\pi\hbar/e$ of
the CS field $C_{\mu}(x)$, with the Lagrangian~\cite{LF,HLR}
(with $e$ restored)
\begin{eqnarray}
L&=& \phi^{\dag} \left[ i\partial_{\,t} 
- {1\over{2M}}{\bf \Pi}[{\bf A}^{\!B}\!+\!{\bf A}\! 
+\!{\bf C}]^{2} -e A_{0} -e C_{0}\right] \phi
\nonumber\\  
&&-{1\over{2}}\,  \delta \rho\, V\, \delta \rho  
-{e\over{2\alpha\,\phi_{D}}}\ C_{\mu}
\epsilon^{\mu\nu\lambda}\partial_{\nu} C_{\lambda}; 
\label{Lphi} 
\end{eqnarray}
${\bf \Pi}[{\bf A}]\equiv {\bf p} +e{\bf A}$.
Here $\delta \rho (x) =\rho (x) -\bar{\rho}$ is written in terms
of $\rho (x)=\phi^{\dag}(x) \phi (x)$, but one can
effectively replace it by 
\begin{equation}
\delta \rho (x) = -(1/\alpha\,\phi_{D})\,
\epsilon^{0ij}\partial_{i}C_{j}(x) -\bar{\rho}.
\label{drhoC}
\end{equation}
To see this, express the Coulomb interaction in linearized
form~(\ref{linearizedCI}) again, and rewrite $L$ in favor of
$C'_{0} = C_{0} + (1/e)\,\chi$. The $C\epsilon \partial C$ term
thereby acquires a
$\chi$-dependent piece, which, upon eliminating $\chi$, 
recovers the Coulomb interaction with 
$\delta \rho (x)$ of Eq.~(\ref{drhoC}). 
With this in mind we now start with Eqs.~(\ref{Lphi}) and
(\ref{drhoC}).

The mean-field treatment corresponds to an expansion in
$C_{\mu}(x)$ of $L$ around the mean field $\delta \rho \sim 0$,
or $\langle C_{j}(x)\rangle$ with 
$\epsilon^{ij}\partial_{i}\langle C_{j}(x)\rangle =
-\alpha\,\phi_{D}\,
\bar{\rho}$ and $\langle C_{0}(x)\rangle =0$.
Let us set $C_{\mu}(x)=\langle C_{\mu}(x)\rangle +c_{\mu}(x)$ 
and rewrite the Lagrangian as
\begin{eqnarray}
&&L= 
\phi^{\dag} \left[ i \partial_{\,t} 
- {1\over{2M}}{\bf \Pi}[{\bf A}^{\rm eff}\!+ {\bf a}]^{2} 
-ea_{0}\right]\phi +L_{\rm CS}[c], \label{Lcf}\\
&&L_{\rm CS}[c]=-{e\over{2\alpha\,\phi_{D}}}\,
 c_{\mu}\epsilon^{\mu\nu\lambda}\partial_{\nu} c_{\lambda} 
+ e\bar{\rho}\,c_{0} 
-{1\over{2}}\, \delta \rho\, V\, \delta \rho ,
\label{Lcs}
\end{eqnarray}
where $a_{\mu}=A_{\mu}+c_{\mu}$ and 
$\delta \rho (x)= -(1/\alpha\,\phi_{D})\,c_{12}(x)$.  

The composite fermions, coupled to  
${\bf A}^{\rm eff}(x) 
= {\bf A}^{\!B}({\bf x}) +\langle {\bf C}(x)\rangle$,
experience a reduced mean magnetic field
$B_{\rm eff}\equiv B-\alpha\,\phi_{D}\bar{\rho} =
(1-\alpha\,\nu)B$, and form Landau levels of smaller gap
$\omega_{\rm eff}=e|B_{\rm eff}|/M$ and state density
$1/(2\pi\ell^{2}_{\rm eff})$, where
$\ell_{\rm eff}\equiv 1/\sqrt{e|B_{\rm eff}|}$. 
The fractional quantum Hall states of interacting electrons
in a magnetic field $B$ at the principal filling fractions 
\begin{equation}
\nu \equiv 2\pi\, \ell^{2}\,\bar{\rho}
=\nu_{\rm eff}/(2p \nu_{\rm eff} \pm1) ,
\label{fqhnu}
\end{equation}
where $\pm$ refers to the sign of $B_{\rm eff}/B$, are thereby
mapped into an integer quantum Hall state of composite fermions
in the reduced field
$B_{\rm eff}$ at integer filling
$\nu_{\rm eff}\equiv 
2\pi \ell^{2}_{\rm eff}\,\bar{\rho}=1,2,\cdots$.  This Jain's 
picture~\cite{J} of (supposedly weakly-interacting) composite
fermions has a number of consequences that have been supported
experimentally.~\cite{expr}

Consider now a many-body ground state of composite fermions at
integer filling  $\nu_{\rm eff}$. Its response to  
$A_{\mu}+c_{\mu}$ is described by 
$L[A_{\mu}+c_{\mu};\nu_{\rm eff}]$, i.e.,
Eq.~(\ref{LAnu}) with
$A_{\mu}\rightarrow A_{\mu}+c_{\mu}$ and
$B \rightarrow B_{\rm eff}$.
Note further that, since the ground state at integer
filling is incompressible and
nondegenerate, it supports vortices as elementary excitations
over it, which as before are introduced via the replacement
$A_{\mu} \rightarrow A_{\mu} + (1/e)\partial_{\mu}\Theta$. 
The composite-fermion (CF) theory~(\ref{Lcf}), generalized to
accommodate vortices, is thus described by the effective
Lagrangian 
\begin{equation}
 L^{\rm CF}_{\rm eff}=L[A_{\mu}+
(1/e)\partial_{\mu}\Theta + c_{\mu};\nu_{\rm eff}] +
L_{\rm CS}[c].
\label{LCFeff}
\end{equation}
This is immediately transcribed into an equivalent 
bosonic version of the CF theory, with the Lagrangian
\begin{eqnarray}
L^{\rm CF}[b,c]= &&-e(A_{\mu} + c_{\mu}) 
\epsilon^{\mu\nu\rho} \partial_{\nu}b_{\rho} 
-2\pi \tilde{j}_{\mu}b_{\mu} 
\nonumber\\
&& + L^{(0)}[b\,; B_{\rm eff}] 
+ L_{\rm CS}[c],
\label{LCFb}
\end{eqnarray}
where $b_{\mu}$ bosonizes the 3-current of the CF field $\phi$;
and $L^{(0)}[b; B_{\rm eff}]$ stands for $L^{(0)}[b]$ of
Eq.~(\ref{Lzerob}) with
$\ell \rightarrow \ell_{\rm eff}, 
\omega \rightarrow \omega_{\rm eff}, 
\nu \rightarrow \nu_{\rm eff}$ and 
$s_{B}\rightarrow s_{B^{*}}$, where $s_{B^{*}}=\pm 1$ refers to
the sign of $B_{\rm eff}/B$.

Let us here try to eliminate $c_{\mu}$ 
from $L^{\rm CF}[b,c]$ and obtain an equivalent theory of
the $b_{\mu}$ field alone.  
With the shift $c_{\mu}=c'_{\mu} -\alpha \phi_{D}\,b_{\mu}$,
the $c \epsilon \partial b$ term and
$L_{\rm CS}[c]$ combine to yield 
\begin{equation}
\triangle L=2\pi \alpha ({1\over{2}}\, b \epsilon
\partial b - \bar{\rho}\,b_{0})
-{1\over{2}}\, \delta b_{12}\, V \, \delta b_{12},
\end{equation}
which is combined with the rest of terms in $L^{\rm CF}[b,c]$
to give the desired $b$-field Lagrangian $L^{\rm (CF)}_{\rm
eff}[b]$. Actually this $L^{\rm (CF)}_{\rm eff}[b]$ almost
coincides with
$L_{\rm eff}[b]$ in Eq.~(\ref{Leffb}) and their apparent
difference lies in 
\begin{eqnarray}
\triangle L +L^{(0)}[b; B_{\rm eff}]\ 
\leftrightarrow \ L^{(0)}[b] .
\label{compare}
\end{eqnarray}
As a matter of fact, both sides agree precisely for the first
three leading terms of $L^{(0)}[b]$; i.e.,
$\alpha\nu /\ell^{2} +s_{B^{*}}/\ell_{\rm eff}^{2}=1/\ell^{2}$
for the $b_{0}$ term, 
$\alpha+ s_{B^{*}}/\nu_{\rm eff}= 1/\nu$
for the $b\epsilon \partial b$ term, and 
$\nu_{\rm eff}\omega_{\rm eff}=\nu \omega$ for the
$(b_{k0})^{2}$ term, with the sign $s_{B^{*}}=\pm 1$ 
properly taken into account. 
Note, however, that 
$\sigma(\nu)/\nu\not= \sigma(\nu_{\rm eff})/\nu_{\rm eff}$ in
general.  Thus there is a discrepancy in the
$(b_{12})^{2}$ term, which fortunately is unimportant at long
wavelengths, as noted for $L_{\rm eff}[b]$.

This leads to two important observations: First, the fact that 
both sides of Eq.~(\ref{compare}) are practically the same shows
the mutual consistency between the fermionic
CS theory and our approach, and makes
it clear again that the effective Lagrangian $L_{\rm eff}[b]$
of Eq.~(\ref{Leffb}) rests on general grounds. The RPA response
of the fermionic CS theory is obtained here from the bosonic
$L^{\rm CF}[b,c]$ theory  by integrations over $b_{\mu}$ and
$c_{\mu}$, with the result given by $L_{b}[A;\nu]$ of
Eq.~(\ref{LArpa}).

Secondly, a tiny discrepancy in the $(b_{12})^{2}$ term
indicates that both sides of Eq.~(\ref{compare}) actually
differ beyond the long-wavelength regime. This discrepancy is
ascribed to a mismatch between the Landau levels of the
original electron and those of the composite fermion, which is
inevitable because the procedures of CS-flux attachment and
Landau-level projection do not commute.  The shorter-wavelength
regime, of course, is beyond the scope of both the fermionic CS
theory and our approach, and it is already nontrivial that both
descriptions are perfectly consistent in the long-wavelength
regime.

In the CF description of the FQHE the composite fermion itself
constitutes an elementary excitation in the FQH ground states.
In particular, Goldhaber and Jain~\cite{GJ} identified a
composite fermion as a dressed electron with bare charge $-e$
and argued by exploiting the incompressible nature of Laughlin's
wave functions that the bare charge is screened in the CF
medium to yield local charge equal to $-\nu e$, consistent with
Laughlin's quasiparticles.

It is possible to substantiate such characteristics of
composite fermions within the CS approach.
As revealed by Laughlin's reasoning, adiabatically piercing the
CF ground state of integer filling $\nu_{\rm eff}$ with a
thin flux of vorticity
$q=-s_{B^{*}}=\mp 1$ (depending on the sign of $B_{\rm eff}/B$)
introduces a hole per filled level and thus $\nu_{\rm eff}$
holes in total.  It is quite obvious at this level of composite
fermions that the quasiholes in Laughlin's picture  are nothing
but the vacancies of the associated composite fermions, and
that the quasielectrons are the composite fermions themselves.
The charges of these excitations are readily read from the
bosonic CF Lagrangian $L^{\rm CF}[b,c]$ in Eq.~(\ref{LCFb}):
As discussed in the previous section, the vortex is coupled to
$A_{\mu}$ through a background piece of
$b_{\mu}$, 
\begin{equation}
-2\pi \tilde{j}_{\mu}b_{\mu}= - \nu_{\rm
eff}\, e\, s_{B^{*}}\,
\tilde{j}_{\mu}(A_{\mu} +c_{\mu})+ \cdots,
\label{vortexcoupl}
\end{equation}
which indicates that each composite fermion 
has ``bare'' charge $-e$ in  response to $A_{\mu} +c_{\mu}$.

Note next that, upon integration over $b$, $L^{\rm CF}[b,c]$
gives rise to a CS term of the form
$-{1\over{2}}\beta_{\phi}\, (A+c) \epsilon \partial (A+c)$
with
$\beta_{\phi}=s_{B^{*}}\,\nu_{\rm eff}\,e^2/(2\pi)$, as 
seen also from $L^{\rm CF}_{\rm eff}$ in Eq.~(\ref{LCFeff}). 
This term combines with another CS term  in $L_{\rm CS}[c]$, 
$-{1\over{2}}\beta_{c}\, c \epsilon \partial c$ with
$\beta_{c}=e^2/(2\pi\alpha)$, to fix the background piece of
$c_{\mu}$ as $c_{\mu}^{\rm bg}= -\{\beta_{\phi}/(\beta_{\phi}
+\beta_{c})\} A_{\mu} + O(\partial A)$. As a result, the
dressing (or renormalization) of the vortex
coupling~(\ref{vortexcoupl}) by the CS field is substantial:
\begin{equation}
A_{\mu}+c_{\mu} = s_{B^{*}}\,
{\nu\over{\nu_{\rm eff}}}\,A_{\mu} +\cdots,
\label{AplusdC}
\end{equation}
which shows that a composite fermion has
fractional charge 
\begin{equation}
Q_{\rm CF}=-{\nu\over{\nu_{\rm eff}}}\, e 
= -{1\over{2p\nu_{\rm eff} \pm1}}\, e 
\label{chargeCF}
\end{equation}
in response to $A_{\mu}$, where $\pm$ refers to $s_{B^{*}}$. 
Hence the bare charge $-e$ coupled to $A_{\mu}+c_{\mu}$ is the
same as the renormalized charge $Q_{\rm CF}$ probed with
$A_{\mu}$. This special renormalization feature of the linear
$A_{\mu}+c_{\mu}$ coupling is a consequence of the
dynamics in the fermionic CS theory.
With this in mind one can directly learn from the original
Lagrangian~(\ref{Lphi}) that the CF field $\phi$, when
probed with $A_{\mu}$, has fractional charge $Q_{\rm CF}$.

It will be worth remarking here that the coefficient 
$\beta_{\phi}=s_{B^{*}}\,\nu_{\rm eff}\,e^2/(2\pi\hbar)$ of the
CS term mentioned above stands for the ``bare'' Hall conductance
in response to $A_{\mu}+c_{\mu}$ for $\nu_{\rm eff}$ filled CF
Landau levels, and that it is thus transcribed into the
(physical) Hall conductance $\sigma_{xy}=-\nu\,e^2/(2\pi\hbar)$
for the associated FQH state in response to $A_{\mu}$.
In other words, the composite fermion with bare charge
$-e$ feels an effective electric field $E^{A+c}_{y}\equiv
E_{y}+c_{20}=  s_{B^{*}}\, (\nu /\nu_{\rm eff}) E_{y} + \cdots$
in the effective magnetic field
$B_{\rm eff}= s_{B^{*}}\,(\nu /\nu_{\rm eff})B$ so that 
it drifts with the same velocity as the
electron, as it should,
\begin{equation} 
v_{x}^{\rm drift}=E^{A+c}_{y}/B_{\rm eff}=E_{y}/B.
\end{equation}

\section{Summary and discussion}

In this paper we have studied the electromagnetic
characteristics of the FQH states by formulating an
effective vector-field theory that properly takes into account
the Landau-level structure and quenching of the cyclotron
energy. The effective theory has been constructed, via
bosonization, from the electromagnetic response of Hall
electrons, in which, as we have seen, the long-wavelength
characteristics of the FQH states and the associated
quasiparticles are correctly encoded. We have thereby
reproduced the dual-field Lagrangian of Lee and Zhang without
invoking the composite-boson picture.

Our approach does not refer to either the composite-boson or
composite-fermion picture, and simply supposes a FQH ground
state of uniform density (in the absence of an external
probe).  
Our approach by itself does not tell at which filling fraction
$\nu$ such a  state is realized. Instead, it tells us that once
such an incompressible state is formed its long-wavelength
characteristics are fixed universally, independent of the
composite-boson and composite-fermion pictures. In this sense,
our approach is complementary to the CS approaches, both
bosonic and fermionic, where the characteristic filling
fractions are determined from the picture-specific condition
for the emergence of composite-boson condensates or filled
composite-fermion Landau levels. 

All these three approaches are consistent at long wavelengths, as
we have verified. This gives further evidence for the universality of the
long-distance physics of the FQHE, as advocated in Ref.~25.
Remember, however, that they start to deviate
at shorter wavelengths; thus care is needed in studying the
detailed features of the FQH states.

It could happen that a bosonic effective theory better reveals
some basic features of the original fermionic theory. Indeed,
we have derived a vector-field version of the fermionic CS
theory and found a special renormalization pattern of the
electron charge, which substantiates the identification  by
Goldhaber and Jain of the composite fermion as a dressed
electron. Looking into the FQHE from various angles would
promote our understanding of it.

\acknowledgments

This work is supported in part by a Grant-in-Aid for 
Scientific Research from the Ministry of Education of Japan, 
Science and Culture (No. 10640265).

\appendix

\section{Calculation}
In this appendix we outline the construction of the
projected Hamiltonian~(\ref{HGprime}).
Let us begin with sorting the $U(\infty)$
basis in Eq.~(\ref{Uinfinity}) 
\begin{eqnarray}
\gamma_{r}^{k}\equiv {\bar{Z}^{r}Z^{r+k}\over{r!(r+k)!}} ,\ 
\bar{\gamma}_{r}^{k}\equiv (\gamma_{r}^{k})^{\dag},
\end{eqnarray} 
into diagonal components $\gamma_{r}^{0}$ and
off-diagonal components $\gamma_{r}^{k}$ with $k\ge 1$.  
One can expand, e.g., $A(\hat{\bf x},t)$ in this basis as
$A(\hat{x},\hat{y},t)
=e^{\bar{Z}\bar{\partial}}e^{Z\partial}A_{00}(x_{0},y_{0},t)$,
or
\begin{eqnarray}
A(\hat{\bf x},t)
&=&\!\sum_{r=0}^{\infty} \gamma_{r}^{0}\,
A_{00}^{(r)} +\!
\sum_{r=0}^{\infty}\sum_{k=1}^{\infty}
\left(
\gamma_{r}^{k}\,
\partial^{k}A_{00}^{(r)}+
\bar{\gamma}_{r}^{k}\,
\bar{\partial}^{k}A_{00}^{(r)}
\right). \nonumber\\
\end{eqnarray}
Here 
$\partial=(\partial_{y_{0}}-i\partial_{x_{0}})/\sqrt{2}$ and
$\bar{\partial}=(\partial_{y_{0}}+i\partial_{x_{0}})/\sqrt{2}$
act on $A_{00}(x_{0},y_{0},t)$ and $A_{00}^{(r)}\equiv
(\bar{\partial}\partial)^{r} A_{00}(x_{0},y_{0},t)$; we have
set $\ell=1$. 
The expression for $A_{00}(x_{0},y_{0},t)$ depends on how
$x_{0}$ and $y_{0}$ are ordered in it. 
The convention we adopt below is to define
$A(\hat{\bf x},t)$ through the Fourier transform of
$A({\bf x},t)$ with the substitution $e^{i{\bf p\cdot x}}
\rightarrow  e^{i{\bf p\cdot \hat{x}}}$, which yields 
the expression
\begin{equation}
A_{00}(x_{0},y_{0},t)= e^{{1\over2}\bar{\partial}\partial} 
A(x_{0},y_{0},t).
\end{equation}
Actually such reference to a particular ordering
convention chosen disappears when one goes back 
to the ${\bf x}$ space.

The gauge transformation law~(\ref{AG}) or its first-order
form $(A^{G})^{[1]}\equiv A + [\eta, Z]$ tells us that, with a
suitable choice of $G$, $A^{G}$ is brought to have only
$\gamma^{k}_{r}$ components.  This is the key to systematizing
the calculation.  Indeed, to $O(A)$, $A^{G}$ is written in
terms of the field strength $h_{z}$,
\begin{eqnarray}
i(A^{G})^{[1]}
&=& \sum_{r=0}^{\infty} \gamma_{r}^{1}\,  {1\over2}
(h_{z})_{00}^{(r)}
 + \sum_{r=0}^{\infty}\sum_{k=2}^{\infty} 
\gamma_{r}^{k}\, \partial^{k-\!1}
(h_{z})_{00}^{(r)},  
\end{eqnarray}
upon choosing $\eta$ as
\begin{eqnarray}
\eta^{[1]} &=& \eta^{0}_{0}+\sum_{r=1}^{\infty}
\gamma_{r}^{0}\, {1\over2}
(\partial A_{00}^{(r-1)}\!
+\!\bar{\partial}\bar{A}_{00}^{(r-1)} )
\nonumber\\  &&
+  \sum_{r=0}^{\infty}\sum_{k=1}^{\infty} 
\left( \gamma_{r}^{k}\, 
\partial^{k-1} \bar{A}_{00}^{(r)} 
+ \bar{\gamma}_{r}^{k}\, \bar{\partial}^{k-1}
A_{00}^{(r)}\right),
\end{eqnarray}
with $(h_{z})_{00}^{(r)}\equiv
(\bar{\partial}\partial)^{r}(h_{z})_{00}(x_{0},y_{0},t)$, etc.
At the same time the scalar potential 
$(A_{0}^{G})^{[1]} \equiv A_{0} - \partial_{t} \eta$ 
reads
\begin{eqnarray}
(A_{0}^{G})^{(1)}
&=& (A'_{0})_{00} + \sum_{r=1}^{\infty} \gamma_{r}^{0}\,
{1\over2}
(\partial E_{00}^{(r-1)}\!
+\!\bar{\partial}\bar{E}_{00}^{(r-1)} )
\nonumber\\ &&
+\sum_{r=0}^{\infty}
\sum_{k=1}^{\infty}(\gamma_{r}^{k}\,
\partial^{k-1}\bar{E}_{00}^{(r)}+
\bar{\gamma}_{r}^{k}\,
\bar{\partial}^{k-1}E_{00}^{(r)}) ,
\end{eqnarray}
with
$E_{00} = (E_{y} + iE_{x})_{00}/\sqrt{2}
=\bar{\partial}(A_{0})_{00} - \partial_{t}A_{00}$, etc. 
The $(A'_{0})_{00}\equiv (A_{0})_{00} 
-\partial_{t}\eta^{0}_{0}$ takes a gauge-invariant form
$(A'_{0})_{00}=A_{0} + g({1\over2}\bar{\partial}\partial)\,
{1\over4}\, {\bf \nabla\cdot E}$, upon choosing $\eta^{0}_{0}=
g({1\over2}\bar{\partial}\partial)\,{1\over4}\,(\partial
A+\bar{\partial}\bar{A})$, where $g(x)=(e^{x}-1)/x$.

One can go to higher orders in $A_{\mu}$ by fixing
$\eta$ so that $A^{G}$ has only $\gamma\,_{r}^{k}$ components. 
Here we construct $H^{G}$ to $O(A^{2})$ and up to two
derivatives.
A direct calculation yields 
\begin{eqnarray}
i(A^{G})^{[2]}
&=& Z\,\Big\{{3\over8}\, h_{00}^{\,2} -
{1\over2}\partial\bar{\partial}(\bar{A} A) 
+ {1\over2}{\rm Re}\,(\partial^{2} A^{2}) \Big\}  +
O_{3}, \nonumber\\
(A_{0}^{G})^{[2]}
&=& {1\over2}\, [\epsilon^{\mu \nu \lambda}
A_{\mu}\partial_{\nu} A_{\lambda} 
- \epsilon^{0jk}\partial_{j}(A_{k} A_{0})] 
\nonumber\\&& \ \ \ \ \ \ \ \ 
+ Z\,O_{2}+\bar{Z}\,O_{2} + O_{3} ,
\end{eqnarray}
where $O_{n}$ stands for terms with $n$ derivatives.
Finally substitute these $A^{G}, A_{0}^{G}$ and 
$G h_{z} G^{-1} =  h_{z} + (h_{z})_{00}^{2} +{\rm total\
deriv.} +\cdots$ into the transformed Hamiltonian
$\hat{H}^{G}$, and make a further transformation to remove
off-diagonal pieces $\propto Z, \bar{Z}$ from it.
This leads to $\hat{H}^{G}$ in Eq.~(\ref{HGprime}).

\section{Integration over Chern-Simons fields}
In the text we frequently calculate a functional integral over
a 3-vector field $b_{\mu}$, with a Lagrangian of the form
\begin{eqnarray}
L[b]&=&  -A_{\mu} \epsilon^{\mu \nu\rho}
\partial_{\nu}b_{\rho} - {1\over{2}}\,\beta\, b_{\mu}
\epsilon^{\mu\nu\rho}
\partial_{\nu} b_{\rho} 
\nonumber\\&&
+{1\over{2}}\,b_{k0}\Gamma\,b_{k0} 
-  {1\over{2}}\ b_{12}\,\Sigma\ b_{12}  -\kappa\, b_{0},
\label{LbA}
\end{eqnarray}
where $\beta$ and $\kappa$ are real constants; $\Gamma$ and
$\Sigma$ may contain derivatives. Our task is to derive the
response to an external potential $A_{\mu}$.  The relevant
integration is best carried out in the following way:
First shift the field  $b_{\mu}=b'_{\mu} +f_{\mu}$ and choose
$f_{\mu}$ so that no direct coupling between $A_{\mu}$ and 
$b'_{\nu}$ remains, i.e.,
\begin{eqnarray}
&&f_{0}+ (1/\beta)\, \Sigma \, f_{12}
= -  (1/\beta)\, A_{0} ,\nonumber\\
&&f_{j}-(1/\beta)\,\epsilon^{jk}\,\Gamma\,
f_{k0}= - (1/\beta)\, A_{j}.
\end{eqnarray}
This is readily solved for $f_{\mu \nu}$,
\begin{eqnarray}
&&f_{k0}= - {1\over{\beta D}}\,
\{A_{k0} - {1\over{\beta}}\, \epsilon^{kj} \Gamma\,
\partial_{0}A_{j0}  
- {1\over{\beta}}\, \Sigma\,\partial_{k}A_{12}\},
\nonumber\\
&&f_{12}= -{1\over{\beta D}}\, \{ A_{12} 
-{1\over{\beta}}\,\Gamma\, \partial_{k}A_{k0}\},\nonumber\\
&&D=1+ (1/\beta^{2})\, 
(\Gamma^{2}\,\partial_{0}^{2} - \Gamma \Sigma\,\nabla^{2}),
\end{eqnarray}
and for $f_{\mu}$ as well. 

Integration over $b'_{\mu}$ (which requires fixing a gauge)
thereby becomes trivial, yielding no dependence on $A_{\mu}$.
All the dependence on $A_{\mu}$ is now isolated in $f_{\mu}$, 
in terms of which the stationary action or the effective
Lagrangian is given by 
$L_{\rm eff}[A]= -{1\over{2}}\,A \epsilon \partial f 
- \kappa\,f_{0}$, or explicitly 
\begin{eqnarray}
 L_{\rm eff}[A]= 
&&{1\over{2\beta}}\,A{1\over{D}}\,\epsilon \partial A 
 +{1\over{2\beta^{2}}}\, F_{k0}{1\over{D}} \Gamma F_{k0} 
\nonumber\\&&
- {1\over{2\beta^{2}}}\,F_{12} {1\over{D}}\Sigma 
F_{12} +{\kappa \over{\beta}}\,A_{0},
\label{resultL}
\end{eqnarray}
apart from total derivatives.  Remember that the
leading part of the background piece
$f_{\mu}=-(1/\beta)\, A_{\mu} + O(\partial A)$ is fixed from
the first two terms in Eq.~(\ref{LbA});
extensive use of this fact is made in the text.
Note finally that Eq.~(\ref{resultL}) loses sense for $\beta
\rightarrow 0$. This shows that the presence of the CS term
$\propto b\epsilon \partial b$ is crucial for the
present method to work. For $\beta=0$ one has to first fix
a gauge for $b_{\mu}$ and calculate the response to $A_{\mu}$.



\begin{thebibliography}{99}

\bibitem{TSG} 
D. C. Tsui, H. L. Stormer, and A. C. Gossard,  Phys. Rev. Lett.
{\bf 48}, 1559 (1982).

\bibitem{Rev} For a review, see {\em The Quantum Hall Effect}, 
edited by R. E. Prange and S. M. Girvin (Springer-Verlag, Berlin, 1987).

\bibitem{L} R. B. Laughlin, Phys. Rev. Lett. {\bf 50}, 1395
(1983).

\bibitem{H} F. D. M Haldane,  Phys. Rev. Lett. {\bf 51}, 605
(1983).

\bibitem{Halp} B. I. Halperin, Phys. Rev. Lett. {\bf 52}, 1583
(1984).


\bibitem{GM} S. M. Girvin and A. H. MacDonald,
 Phys. Rev. Lett. {\bf 58}, 1252 (1987). 

\bibitem{R} N. Read, Phys. Rev. Lett. {\bf 62}, 86 (1989). 

\bibitem{J} J.~K.~Jain, Phys. Rev. Lett.
{\bf 63}, 199 (1989); Phys. Rev. B {\bf 41}, 7653 (1990). 

\bibitem{ZHK} S. C. Zhang, T.H. Hansson, and S. Kivelson, 
 Phys. Rev. Lett. {\bf 62}, 82 (1989). 

\bibitem{LZ} D.-H. Lee and S.-C. Zhang, 
 Phys. Rev. Lett. {\bf 66}, 1220 (1991). 

\bibitem{Z} S. C. Zhang, 
 Int. J. Mod. Phys. B {\bf 6}, 25 (1992). 

\bibitem{BW} B. Blok and X. G. Wen, Phys. Rev. B {\bf 42}, 8133
(1990).

\bibitem{LF} A. Lopez and E. Fradkin, Phys. Rev. B {\bf 44},
5246 (1991); {\it ibid.}  {\bf 47}, 7080 (1993).

\bibitem{HLR} B.~I.~Halperin, P.~A.~Lee and N.~Read, 
Phys. Rev. B {\bf 47}, 7312 (1993). 

\bibitem{KSw} K. Shizuya, Phys. Rev. B {\bf 45}, 11 143 (1992);
{\it ibid.} {\bf 52}, 2747 (1995). 


\bibitem{sakita} P. K. Panigrahi, R. Ray, and B. Sakita,
Phys. Rev. B {\bf 42}, 4036 (1990).

\bibitem{FS} E. Fradkin and F. A. Schaposnik, Phys. Lett. B
{\bf 338}, 253 (1995).

\bibitem{schaposnik} F. A. Schaposnik, Phys. Lett. B {\bf 356},
39 (1995).

\bibitem{BOS} D. G. Barci, C. A. Linhares, A. F. De Queiroz,
and J. F. Medeiros Neto, Int. J. Mod. Phys. A {\bf 15}, 4655 
(2000). 

\bibitem{BO} D. G. Barci and L. E. Oxman, Nucl. Phys. B {\bf 580}, 721
(2000).  


\bibitem{LeeF} D.-H. Lee and M.~P.~A. Fisher, Phys. Rev. Lett.
{\bf 63}, 903 (1989).

\bibitem{K} W. Kohn, Phys. Rev. {\bf 123}, 1242 (1961).

\bibitem{expr} 
R. L. Willett, R. R. Ruel, K. W. West, and L. N. Pfeiffer, 
Phys. Rev. Lett. {\bf 71}, 3846 (1993); 
W. Kang, H. L. Stormer, L. N. Pfeiffer, K. W. Baldwin, and K.
W. West,  Phys. Rev. Lett. {\bf 71}, 3850 (1993); 
V. J. Goldman, B. Su, and J. K. Jain,
Phys. Rev. Lett. {\bf 72}, 2065 (1994);


\bibitem{GJ} A.~S.~Goldhaber and J.~K.~Jain, Phys. Lett. A
{\bf 199}, 267 (1995). 

\bibitem{LF} A. Lopez and E. Fradkin, Phys. Rev. Lett. {\bf 69}, 2126
(1992).  


\end{thebibliography}
\end{document}